\documentclass{PoS}

\title{Searches for neutrinos from fast radio bursts with IceCube}

\ShortTitle{Searches for neutrinos from FRBs}

\author{
The IceCube Collaboration\footnote{For collaboration list, see PoS(ICRC2019) 1177.}\\
{\itshape \href{http://icecube.wisc.edu/collaboration/authors/icrc19_icecube}{http://icecube.wisc.edu/collaboration/authors/icrc19\_icecube}}\\
E-mail: \email{akheirandish@icecube.wisc.edu, apizzuto@icecube.wisc.edu, justin.vandenbroucke@wisc.edu}
}

\abstract{

Although IceCube has discovered a diffuse astrophysical neutrino flux, the underlying sources of these neutrinos remain unknown. Transient astrophysical objects, such as fast radio bursts (FRBs), could explain a large percentage of the measured flux. We present the analysis techniques of IceCube searches for MeV to TeV neutrinos from FRBs. As no significant correlation between IceCube neutrinos and FRBs has been found, we present the first limit on MeV neutrino emission from FRBs and the most constraining limits for neutrinos with GeV to TeV energies. We also describe the prospects for future IceCube neutrino searches coinciding with FRB detections from next generation radio interferometers. \\

\vspace{4mm}
{\bfseries Corresponding authors:}
Ali Kheirandish$^{1}$, \speaker{Alex Pizzuto}$^{1}$, Justin Vandenbroucke$^{1}$\\
{$^{1}$ \itshape Department of Physics and Wisconsin IceCube Particle Astrophysics Center, University of Wisconsin, Madison, WI 53706, USA}\\

}

\FullConference{36th International Cosmic Ray Conference -ICRC2019-\\
		July 24th - August 1st, 2019\\
		Madison, WI, U.S.A.}

\begin{document}

\section{Introduction}\label{sec:intro}
The last few years have ushered in the era of multi-messenger astronomy, with evidence for neutrino emission from the blazar TXS0506+056 \cite{IceCube:2018dnn, IceCube:2018cha} as well as the joint detection of GW170817 by LIGO together with observatories across the electromagnetic (EM) spectrum \cite{ANTARES:2017bia}. Despite the evidence for neutrino emission from TXS0506+056, the origins of the overwhelming majority of the measured diffuse astrophysical neutrino flux remain yet unknown \cite{Aartsen:2015knd, Aartsen:2015rwa}. Untriggered searches for neutrino point sources integrating over IceCube's livetime have resulted in no significant detections \cite{Aartsen:2018ywr}, which may be an indication that the majority of neutrino point sources are either very weak or have underlying temporal structure. However, a search for spatially and temporally coincident multiplets in IceCube data suggests that if a single class of transients is contributing to a majority of the diffuse flux, then individual sources in the class must be dim, but the source rate must be sufficiently high to produce the requisite flux \cite{Aartsen:2018fpd}.

Fast radio bursts (FRBs) are a newly discovered and enigmatic class of astrophysical transients characterized by non-periodic millisecond-scale radio flashes. Their high dispersion measures (DM) \cite{Lorimer:2018rwi, Keane:2018jqo} and prevalence at high Galactic latitudes suggest that FRBs are likely of extragalactic origin. To date, a total of 75 FRBs have been detected, but the estimated all-sky rate of detectable FRBs is on the order of thousands per day \cite{Callister:2016vtl}, and this disparity is driven in large part by the narrow field of view of most radio telescopes. If in the future, joint detections of FRBs in radio as well as other EM bands allow for accurate distance measurements, we may find that their redshift distribution and bolometric luminosities make them a prime candidate for explaining the diffuse neutrino flux, and current estimates on rates are consistent with IceCube lower bounds. In addition to energetics and rate requirements, FRB progenitors would need to be sites of hadronic acceleration in order to produce neutrinos. Although most models of FRBs only explain leptonic emission \cite{Platts:2018hiy}, hadronic acceleration is also possible in the associated regions of the progenitors, and the nature of FRB progenitors is still under heated debate. 

The IceCube Neutrino Observatory is a detector consisting of 5160 digital optical modules (DOMs) throughout a cubic kilometer of Antarctic ice between depths of 1450 m and 2450 m. Neutrino interactions within the ice or underlying bedrock lead to the production of charged particles, whose Cherenkov photons are detectable by the DOMs, and which allow for reconstruction of the neutrino energy, direction, and interaction type \cite{Aartsen:2016nxy}. IceCube reached its full detector configuration in 2010, and boasts an up-time higher than 99\%, enabling real-time alerts for other instruments and making it a prime tool to follow-up interesting detections, such as FRBs, from other observatories. Analyses using IceCube data have previously set limits on neutrino emission from FRBs using high-energy neutrino candidate events \cite{Fahey:2016czk, FRB_6yr}. The analyses presented here improve the constraints at these energies and provide a test at a wider range of neutrino energies, which can be used to tailor models of FRB progenitors in the future. 

\section{Analysis Methods}\label{sec:methods}
\begin{figure}
\centering
  \begin{minipage}{0.31\textwidth}
  \centering
  \includegraphics[width=\textwidth]{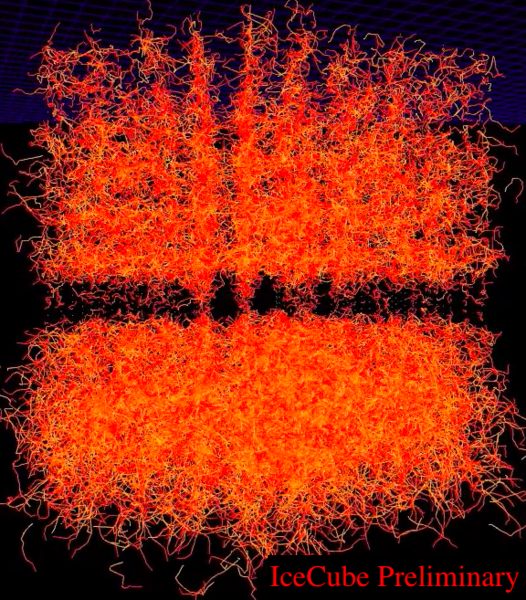}
  \end{minipage}
  \hfill
  \begin{minipage}{0.31\textwidth}  
  \centering 
  \includegraphics[width=\textwidth]{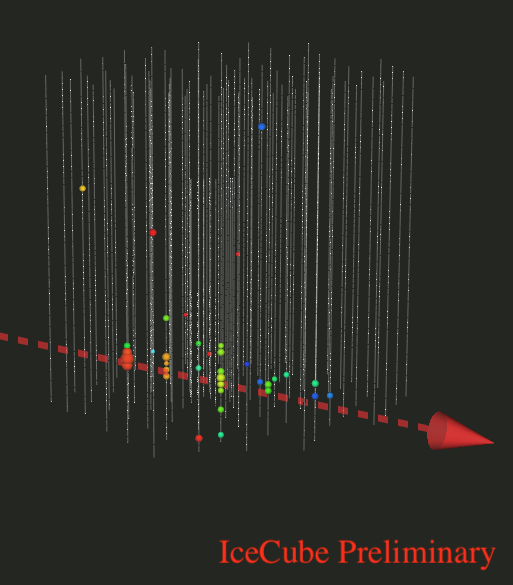}
  \end{minipage}
  \hfill
  \begin{minipage}{0.29\textwidth}  
  \centering 
  \includegraphics[width=\textwidth]{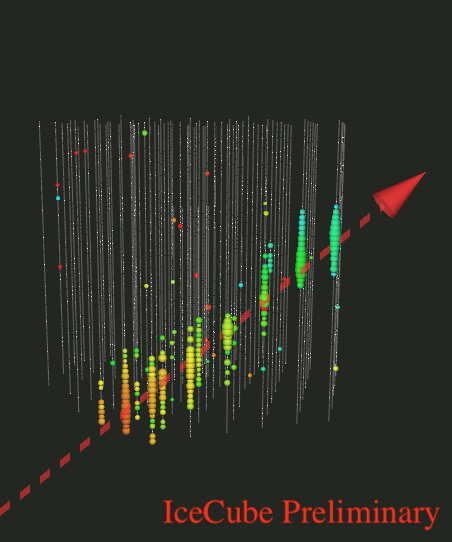}
  \end{minipage}
  \vspace{0.2in}
  \caption{Simulation of the various types of IceCube events analyzed in these analyses. Large fluxes of MeV neutrinos can lead to short tracks that interact with at most one DOM (left). At higher energies, individual charged-current interactions from incident muon-neutrinos lead to long tracks that can be reconstructed. The color scale indicates timing, with early hits in red and later hits in blue. The right shows an example of the high-energy events used in the six-year analysis \cite{FRB_6yr}, while the middle shows an example of the events that are considered in the muon track analysis as described in the text.}
  \label{fig:det}
\end{figure}

 At high energies (greater than  100 GeV), individual charged-current muon neutrino interactions create muons which travel long path lengths through the ice. These tracks allow for per-event reconstructions, where directional resolution increases with energy. In section \ref{subsec:l2}, we discuss how these tracks can be used to search for both temporal and spatial coincidence between neutrinos and FRBs. Additionally, lower energy neutrinos can produce short tracks in the ice. Although each such individual neutrino interaction has a signature which is too small to reconstruct, a large flux of these interactions can give rise to a collective increase in the rate of hits in the detector \cite{Abbasi:2011ss}, and thus if FRBs create a large flux of low energy ($\mathcal{O}$(10 MeV)) neutrinos, then they could be detected by searching for temporal coincidence between FRBs and IceCube hit rates, which is what is presented in \ref{subsec:SNDAQ}. The difference between the two event signatures is displayed in Figure \ref{fig:det}. 
 
\subsection{Search for muon track events}\label{subsec:l2}
\begin{figure}
\centering
  \begin{minipage}{0.48\textwidth}
  \centering
  \includegraphics[width=\textwidth, trim={1cm 1cm 0cm 0cm}]{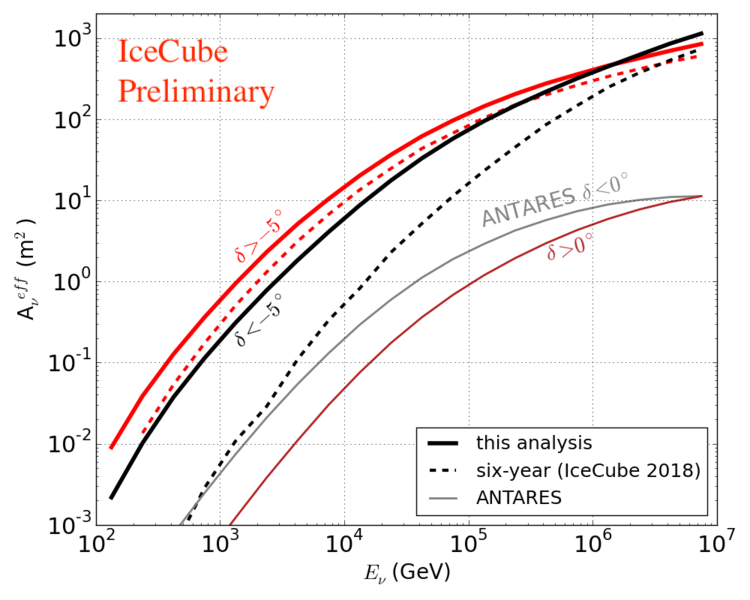}
  \caption{The effective area of this selection to muon neutrinos is significantly improved compared to the six-year analysis, especially in the Southern sky (black). This also improves upon the smaller effective area of ANTARES. Here, we compare to a point-source event selection \cite{ANTARES1} that approximately reproduces the effective area from their 2018 FRB analysis (Figure 5 in \cite{ANTARES2}).}
  \label{fig:eff_area}
  \end{minipage}
  \hfill
  \begin{minipage}{0.48\textwidth}  
  \centering 
  \includegraphics[width=\textwidth, trim={1cm 1cm 1cm 2cm}]{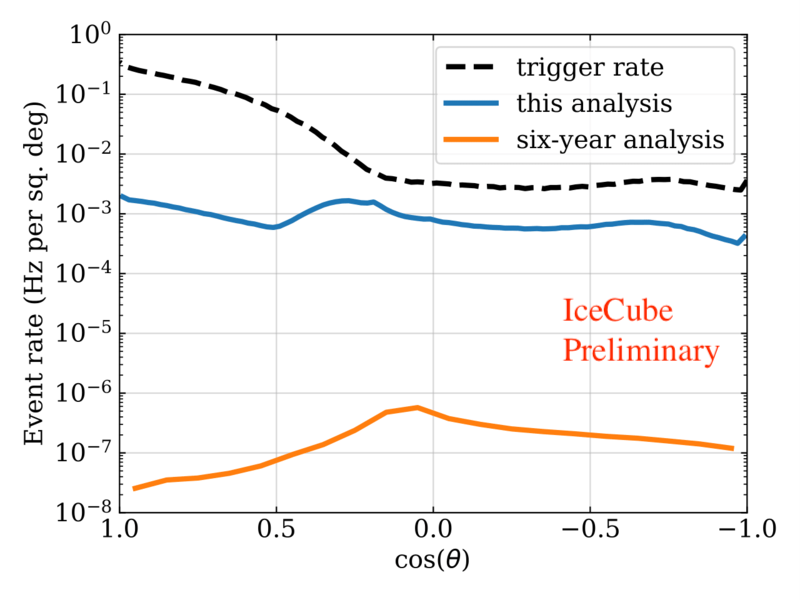}
  \caption{The event rate for these data is significantly larger than that of the six-year analysis, due mostly to an increase in the number of muons produced in the atmosphere passing the event selection. The increase is largest in the Southern sky (${\cos(\theta)>0}$), and is nearly a factor of $10^5$ larger at zenith. Features in the Southern sky are the result of selection methods.}
  \label{fig:zenith_pdf}
  \end{minipage}
  \vspace{0.2in}
\end{figure}

The first analysis is an unbinned maximum likelihood analysis similar to that of IceCube's previous search for neutrino emission from FRBs \cite{FRB_6yr}, hereafter referred to as the \textit{six-year analysis}. Both analyses search for spatial and temporal correlation between neutrinos and FRBs, however, the two analyses differ in their event samples. The six-year analysis used a high-purity track selection, consisting of mainly very high-energy tracks ($E_{\nu} > 10$ TeV) or tracks that had penetrated many kilometers of earth prior to detection, to decrease the probability that an event was atmospheric as opposed to astrophysical in origin. However, FRBs characteristic timescales allow us to loosen the event selection, increasing the effective area at the cost of increased atmospheric backgrounds. For short timescales ($\mathcal{O}$(ms)), this increase in background is subdominant to the gain in effective area in terms of analysis sensitivity. The loosening of the event selection has the largest effect in the Southern Hemisphere for neutrino energies less than around 100 TeV, where the cuts in the six-year analysis are the most severe, as shown in Figures \ref{fig:eff_area} and \ref{fig:zenith_pdf}.

The test statistic (TS) for this analysis is identical to that of the six-year analysis \cite{FRB_6yr}. For a search time window, $\Delta T$, the analysis considers all events within ${[t_{FRB}-\Delta T /2,~~ t_{FRB}+\Delta T /2]}$ to be temporally coincident with the FRB in question. With the lack of a conclusive model for FRBs, no energy dependence is tested in the likelihood, and the TS is given by 

\begin{equation}
\textrm{TS}= -{\hat{n}_s}+\sum_{i=1}^{N}\ln\Big[1+\frac{{\hat{n}_s}S(x_i)}{n_b B(x_i)}\Big] \; .
\label{TS_def}
\end{equation}
The TS is maximized with respect to find the best fit number of observed signal events, $\hat{n}_s$, for events with directional properties $x_i$, on top of the expected number of background events, $n_b$. $S(x_i)$ describes the signal spatial probability distribution function (PDF) that considers the angular distance of an event direction $x_i$ with respect to the coordinates of a given FRB, modeled as a 2D Gaussian, and $B(x_i)$ describes the Poissonian background PDF in that direction, which is parameterized from data (excluding the search time window). 

Two different tests are performed in this analysis, each over a range of $\Delta T$. The first is the stacking test, which tests the hypothesis that the astrophysical class of FRBs emits neutrinos, and simultaneously evaluates the TS for all considered sources. The second, the max-burst test, evaluates a TS separately for each FRB and its respective events, returning only the largest TS as the observation at ${\Delta T}$. This tests the hypothesis that one or a few bright sources emit neutrinos among a heterogeneous class of FRBs. These tests are applied to 28 FRBs. The repeating FRB, FRB 121102, was excluded because all published bursts were included in the six-year analysis, and for a fixed number of bursts, the improvement in sensitivity from the increased effective area is negated by the corresponding enhancement in background rate at its location in the Northern Sky. 

\subsection{Search for temporal MeV neutrino coincidence}\label{subsec:SNDAQ}
In addition to searching for individual tracks from high-energy neutrino interactions, IceCube is also sensitive to the collective increase in photomultiplier tube (PMT) rates on top of backgrond noise. Although this search technique was initially used to search for anti-electron-neutrinos produced by nearby supernovae \cite{Abbasi:2011ss}, it can be used to probe any model that includes large fluxes of $\mathcal{O}$(10 MeV) neutrinos on short-timescales, which is compatible with searching for neutrinos from FRBs. This stream, the Supernova Data Acquisition (SNDAQ) \cite{Abbasi:2011ss}, stores PMT signals, also called DOM \textit{hits}, into bins of 1.6 ms, allowing searches on timescales as small as 2 ms. Here, this capability is used to search for neutrino emission from 21 different FRBs for which data were available (separate bursts from FRB 121102 are classified as distinct FRBs for this analysis). For each FRB, 8 different time windows are analyzed, expanding by powers of 2 from 10 ms up to 1280 ms.

Here, a one-dimensional Gaussian likelihood is used to determine the significance of a collective deviation $(\Delta\mu)$ of the hit rate across the detector in order to find an excess in the DOM rates on top of the background. This likelihood is given by 

\begin{equation}
	\mathcal{L}(\Delta\mu) = \prod_{i=1}^{N_\mathrm{DOM}} \, \frac{1}{\sqrt{2\pi}\,\langle\sigma_i\rangle} \, {\rm exp}(-\frac{(n_i-(\mu_i+\epsilon_i\,\Delta\mu))^2}{2\langle\sigma_i\rangle^2}) \; .
\end{equation}

Here, $n_i$ is the hit rate in DOM$_i$ in a chosen time bin, $\epsilon_i$ is a DOM-specific efficiency parameter that accounts for module and depth dependent detection probabilities, and $\mu_i$ and $\sigma_i$ are the mean and standard deviation for the hit rate in DOM$_i$. The log-likelihood is then maximized with respect to $\Delta\mu$, leading to a significance, $\xi$, given by

\begin{equation}
    \xi = \frac{\Delta\mu}{\sigma_{\Delta\mu}}.
\end{equation}

To quantify the statistical significance of $\xi$, the result is compared to the distribution of significances from off-time windows before and after the run that include the FRB being analyzed. Additionally, it has been shown that the rate of the hits in SNDAQ contains a contribution that is correlated with the seasonally dependent rate of atmospheric muons in the detector. To remove this dependency, we employ the technique used in \cite{Aartsen:2015cwa}. If the resulting statistical significance of the FRB time window is greater than the one-sided $3\sigma$ threshold from seasonally corrected background significances, then we claim detection, otherwise, we report upper limits.

\section{Results}
\label{sec:results}
\begin{figure}[t!]
\centering
  \begin{minipage}{0.48\textwidth}
  \centering
  \includegraphics[width=\textwidth, trim={1cm 3cm 1cm 1cm}]{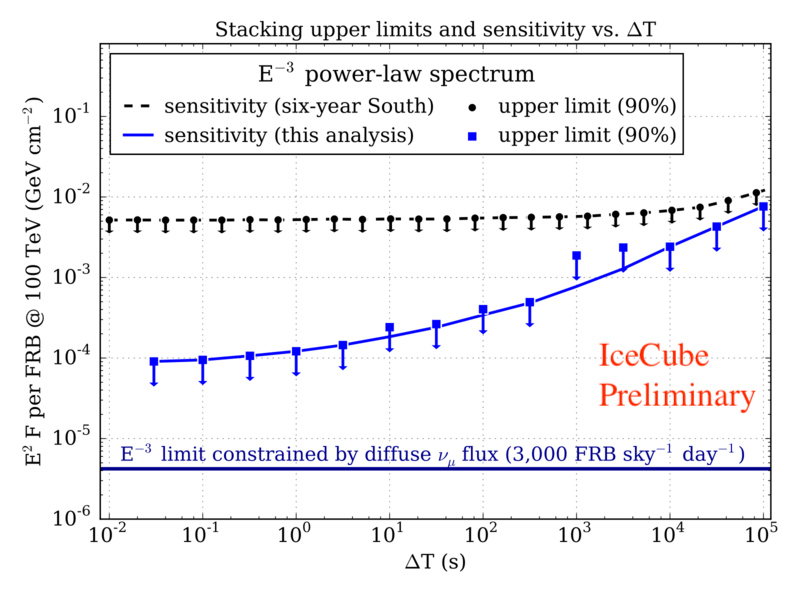}
  \end{minipage}
  \hfill
  \begin{minipage}{0.48\textwidth}  
  \centering 
  \includegraphics[width=\textwidth, trim={1cm 3cm 1cm 1cm}]{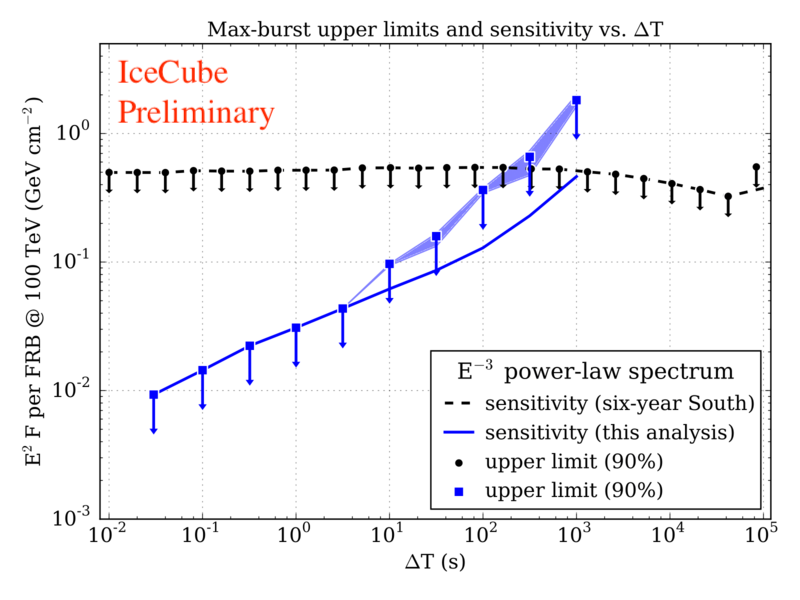}
  \end{minipage}
  \vspace{0.2in}
  \caption{Upper limits on the time-integrated neutrino flux per FRB for a range of $\Delta T$, assuming power-law spectra of $E^{-3}$ (left). The blue line in the stacking limits is produced by dividing IceCube's entire astrophysical $\nu_{\mu}$ flux \cite{Aartsen:2017mau} among a homogeneous class of 3,000 FRBs per day \cite{Callister:2016vtl}. We also constrain the maximum time-integrated neutrino flux among 28 FRBs, assuming the same spectrum (right). The error bands on these limits represents the central 90\% of systematic variation in limits due to uncertainty in background parameterization.}
  \label{fig:limits}
\end{figure}

\begin{figure} [htb!]
\centering
  \makebox[\textwidth]{\makebox[1.05\textwidth]{
  \begin{minipage}{0.4\textwidth}
  \centering
  \includegraphics[width=\textwidth, trim = {0cm 1cm 0cm 0cm}]{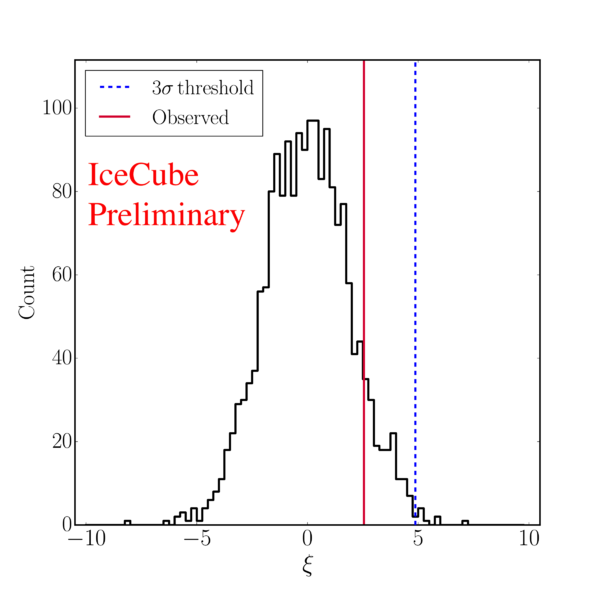}
  \caption{The observed significance (red) for the most significant FRB in this search is compared to the background distribution of the significances. The significance is compared to the 3$\sigma$ threshold (dashed) obtained from off-time period before and after each FRB.}
  \label{fig:sig_distribution}
  \end{minipage}
  \hfill
  \begin{minipage}{0.6\textwidth}  
  \centering 
  \includegraphics[width=\textwidth, trim = {2cm 1cm 2cm 3cm}]{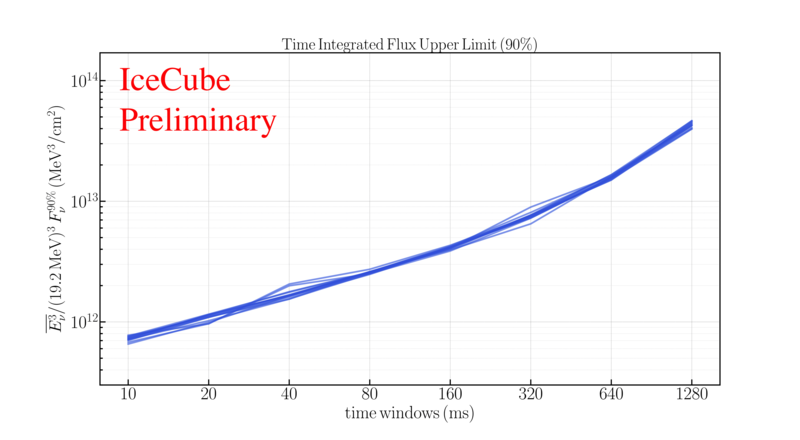}
  \caption{Upper limits (90\% C.L.) on the time-integrated flux of anti-electron neutrinos from FRBs. Each line represents a limit for an individual FRB of the 21 FRBs considered in the analysis, assuming the fiducial model for neutrino emission with a mean neutrino energies of 15.6 MeV.}
  \label{fig:upperlimits}
  \end{minipage}}}
  \vspace{0.2in}
\end{figure}

For both analyses, all results are consistent with expectations from atmospheric background. For the muon track analysis, the smaller time windows are contained within the larger time windows and thus test results from consecutive time windows are correlated in this analysis. To properly calculate the probability of exceeding the smallest pre-trial $p$-value over the course of expansion of $\Delta T$ (find the post-trial $p$-value), we use a Monte Carlo simulation.
Post-trials, the $p$-values for the stacking and max-burst tests are 0.35 and 0.33, respectively. Upper limits are calculated (90\% confidence level) for the time-integrated flux per FRB for each $\Delta T$ (Figure~\ref{fig:limits}). In the stacking search, the limits we set for ${\Delta T < 1 \mathrm{~s}}$ are a factor of 50 stronger on spectra of $E^{-3}$ when compared to the six-year analysis Southern Sky results. In the max-burst search, the same scale of improvement is made on the maximum flux among 28 sources at the smallest $\Delta T$. Limits for an $E^-{2}$ spectrum have also been calculated, and represent an order-of-magnitude improvement for the shortest timescales when compared against the six-year analysis.

In the search for MeV neutrino emission from FRBs, no significance was found above the $3\sigma$ threshold. The distribution of significances from background along with the observed result from the most statistically significant search (2.55 $\sigma$) is shown in Figure \ref{fig:sig_distribution}, along with upper limits on the flux of anti-electron neutrinos for each burst in Figure \ref{fig:upperlimits}. In the absence of neutrino spectrum models for FRBs, the spectrum for neutrino emission from core collapse supernovae is used as a fiducial model, with normalization chosen such that the neutrino flux has average neutrino energy $E_\nu=15.6$ MeV and pinching parameter $\alpha=3$ \cite{Totani:1997vj}, yielding  $<E_\nu^3> = 7118\, \rm MeV^3$, though it is worth noting there is dependency on the signal hit rate from varying incident fluxes \cite{Abbasi:2011ss}. Limits correspond to the 90\% one-sided confidence level for the Gaussian distribution of the significance from the off-time runs. 

\section{Conclusion and Outlook}
In two complementary searches for neutrino emission from FRBs, no significant association has been found. The limits set as a result of these analyses are the most constraining to date. 

\begin{figure}
    \centering
    \includegraphics[width=0.6\textwidth]{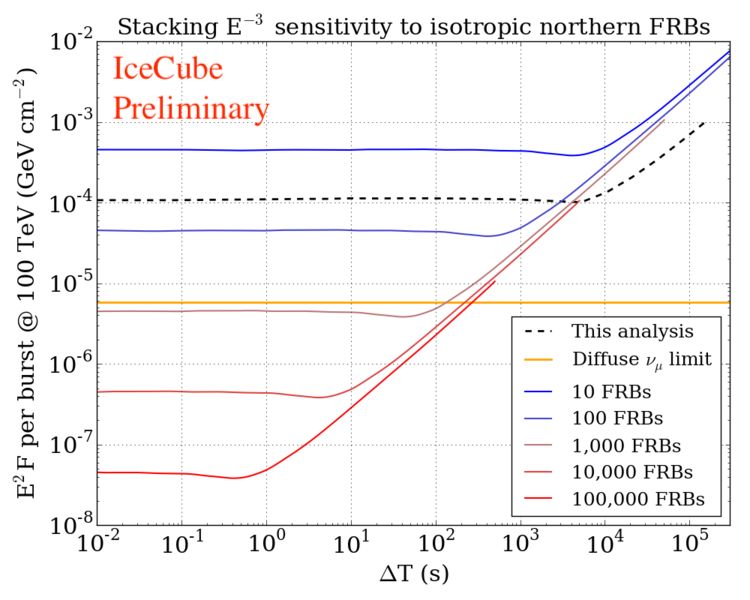}
    \caption{Stacking sensitivity projections for an isotropically distributed class of FRBs, with number of radio detections indicated in the legend. For FRBs in the Northern Sky, sensitivity to a class of FRBs with an $E^{-3}$ spectrum will surpass the limit set by the diffuse $\nu_{\mu}$ flux when the source class increases by one order of magnitude.  }
    \label{fig:Nbursts}
\end{figure}

For the track search, the sensitivity of the analysis scales with the number of detected FRBs. Therefore, as the source class grows, the techniques presented here will be able to probe even smaller per-burst neutrino emission models for small time windows. This effect is amplified by the fact that IceCube's best sensitivity is in the Northern Sky, where interferometers like CHIME \cite{Amiri:2018qsq} will be sensitive. With fields of view drastically larger than single-dish radio telescopes, new radio interferometers are projected to detect on the order of one FRB per day \cite{Amiri:2018qsq, Newburgh:2016mwi}. Projections for the sensitivity of a track based analysis for varying number of FRB detections are shown in Figure \ref{fig:Nbursts}.

The MeV neutrino search was possible because of IceCube's ability to identify bursts of MeV neutrinos from collective increases in DOM hit rates. This offers a unique opportunity for temporal study of low-energy neutrino emission from transients. With this opportunity, we have placed the first ever limits on neutrino signals at MeV energies from FRBs.

\bibliographystyle{ICRC}
\bibliography{references}

\providecommand{\href}[2]{#2}\begingroup\raggedright\begin{thebibliography}{10}

\bibitem{IceCube:2018dnn}
{\bf IceCube, Fermi-LAT, MAGIC, AGILE, ASAS-SN, HAWC, H.E.S.S., INTEGRAL,
  Kanata, Kiso, Kapteyn, Liverpool Telescope, Subaru, Swift NuSTAR, VERITAS,
  VLA/17B-403} Collaboration, M.~G. Aartsen et~al., {\em Science} {\bf 361}
  (2018) eaat1378.

\bibitem{IceCube:2018cha}
{\bf IceCube} Collaboration, M.~G. Aartsen et~al., {\em Science} {\bf 361}
  (2018) 147--151.

\bibitem{ANTARES:2017bia}
{\bf ANTARES, IceCube, Pierre Auger, LIGO Scientific, Virgo} Collaboration,
  A.~Albert et~al., {\em Astrophys. J.} {\bf 850} (2017) L35.

\bibitem{Aartsen:2015knd}
{\bf IceCube} Collaboration, M.~G. Aartsen et~al., {\em Astrophys. J.} {\bf
  809} (2015) 98.

\bibitem{Aartsen:2015rwa}
{\bf IceCube} Collaboration, M.~G. Aartsen et~al., {\em Phys. Rev. Lett.} {\bf
  115} (2015) 081102.

\bibitem{Aartsen:2018ywr}
{\bf IceCube} Collaboration, M.~G. Aartsen et~al., {\em Eur. Phys. J.} {\bf
  C79} (2019) 234.

\bibitem{Aartsen:2018fpd}
{\bf IceCube} Collaboration, M.~G. Aartsen et~al., {\em Submitted to: Phys.
  Rev. Lett.} (2018).

\bibitem{Lorimer:2018rwi}
D.~R. Lorimer, \href{http://arxiv.org/abs/1811.00195}{{\tt arXiv:1811.00195}}.

\bibitem{Keane:2018jqo}
E.~F. Keane, \href{http://arxiv.org/abs/1811.00899}{{\tt arXiv:1811.00899}}.

\bibitem{Callister:2016vtl}
T.~Callister, J.~Kanner, and A.~Weinstein, {\em Astrophys. J.} {\bf 825} (2016)
  L12.

\bibitem{Platts:2018hiy}
E.~Platts, A.~Weltman, A.~Walters, S.~P. Tendulkar, J.~E.~B. Gordin, and
  S.~Kandhai, \href{http://arxiv.org/abs/1810.05836}{{\tt arXiv:1810.05836}}.

\bibitem{Aartsen:2016nxy}
{\bf IceCube} Collaboration, M.~G. Aartsen et~al., {\em JINST} {\bf 12} (2017)
  P03012.

\bibitem{Fahey:2016czk}
S.~Fahey, A.~Kheirandish, J.~Vandenbroucke, and D.~Xu, {\em Astrophys. J.} {\bf
  845} (2017) 14.

\bibitem{FRB_6yr}
{\bf IceCube} Collaboration, M.~G. Aartsen et~al., {\em Astrophys. J.} {\bf
  857} (2018) 117.

\bibitem{Abbasi:2011ss}
{\bf IceCube} Collaboration, R.~Abbasi et~al., {\em Astron. Astrophys.} {\bf
  535} (2011) A109. [Erratum: Astron. Astrophys.563,C1(2014)].

\bibitem{ANTARES1}
{\bf ANTARES} Collaboration, A.~Albert et~al., {\em Phys. Rev.} {\bf D96}
  (2017) 082001.

\bibitem{ANTARES2}
{\bf ANTARES} Collaboration, A.~Albert et~al.,
  \href{http://arxiv.org/abs/1807.04045}{{\tt arXiv:1807.04045}}.

\bibitem{Aartsen:2015cwa}
{\bf IceCube} Collaboration, M.~G. Aartsen et~al., {\it {The IceCube Neutrino
  Observatory - Contributions to ICRC 2015 Part V: Neutrino Oscillations and
  Supernova Searches}},  2015.
\newblock \href{http://arxiv.org/abs/1510.05227}{{\tt arXiv:1510.05227}}.

\bibitem{Aartsen:2017mau}
{\bf IceCube} Collaboration, M.~G. Aartsen et~al.,
  \href{http://arxiv.org/abs/1710.01191}{{\tt arXiv:1710.01191}}.

\bibitem{Totani:1997vj}
T.~Totani, K.~Sato, H.~E. Dalhed, and J.~R. Wilson, {\em Astrophys. J.} {\bf
  496} (1998) 216--225.

\bibitem{Amiri:2018qsq}
{\bf CHIME/FRB} Collaboration, M.~Amiri et~al.,
  \href{http://arxiv.org/abs/1803.11235}{{\tt arXiv:1803.11235}}.

\bibitem{Newburgh:2016mwi}
L.~B. Newburgh et~al., {\em Proc. SPIE Int. Soc. Opt. Eng.} {\bf 9906} (2016)
  99065X.

\end{thebibliography}\endgroup

\end{document}